\documentclass[twoside]{ilcws07}
\usepackage[latin1]{inputenc}
\usepackage[dvips]{graphicx,epsfig,color}
\usepackage{amssymb,amsmath,array}
\begin{document}
\title{
Dark Matter in the U(1) Extended SUSY} 
\author{D. Jarecka$^1$, J. Kalinowski$^1$\thanks{Presented at the International
Linear Collider Workshop 2007: LCWS2007
and ILC2007,  DESY, Hamburg. }, \, S.F. King$^2$ and J.P. Roberts$^1$
\vspace{.3cm}\\
1- Physics Department, University of Warsaw \\
Hoza 69, 00-681 Warsaw, Poland
\vspace{.1cm}\\
2- Department of Physics and Astronomy, University of Southampton \\
Highfield, Souhtampton, SO17 1BJ, United Kingdom \\
}

\maketitle

\begin{abstract}
The  neutralino sector of the U(1)
extended SUSY is presented and some collider and cosmology-related phenomenology discussed.
\end{abstract}

\section{Introduction}
Supersymmetry (SUSY) has been  considered to be the best candidate beyond the standard
model (SM) from a viewpoint of both the hierarchy problem and the gauge coupling
unification.  Recent astrophysical observations  showing the existence of a substantial
amount of non-relativistic and non-baryonic dark matter seem to make SUSY even more promising.
The lightest SUSY particle (LSP) of  R-parity conserving models,
in most cases the lightest neutralino,  can serve as  a good candidate for Dark Matter (DM).

The parameter space of the constrained 
MSSM,   however, is strongly restricted by the requirement of matching the precise
measurement of the DM relic density 
as measured by the WMAP. The MSSM also suffers from a naturalness problem (the
so-called $\mu$ problem): why the dimensionful parameter $\mu$ of the supersymmetric
Higgs mass term $\mu \hat H_1\hat H_2$ has to be of EW scale. This problem can be solved
in the next-to-MSSM (NMSSM) by promoting the $\mu$ parameter to a new  singlet
superfield $S$ coupled to Higgs doublets, $\lambda \hat S\hat H_1\hat H_2$~\cite{nmssm}.
This triple-Higgs coupling term also helps to push up the mass of the lightest CP-even
Higgs boson, relaxing the fine-tuning necessary to comply with the LEP bounds.
Postulating an additional U$_X$(1) gauge symmetry~\cite{ussm} avoids a massless axion, or domain
wall problems of the NMSSM. Such a U(1)-extended MSSM (USSM) can be
considered as an effective low-energy approximation of a more complete E$_6$SSM
model~\cite{King:2005jy}, with other E$_6$SSM fields assumed heavy.

In addition to the MSSM superfields, the USSM contains a chiral
superfield $\hat S$ and an Abelian gauge superfield $B'$. Thus the
MSSM particle spectrum is extended by a new CP-even Higgs boson $S$, a
gauge bozon $Z'$ and two neutral --inos: a singlino $\tilde{S}$ and a
bino' $\tilde{B}'$; other sectors are not enlarged. As a result the
phenomenology of the neutralino sector can be significantly modified
both at colliders~\cite{Choi:2006fz} and in cosmology-related
processes~\cite{Barger:2007nv,jkkr}.  To illustrate this we consider a physically
interesting scenario with higgsino and gaugino mass parameters of the
order $M_{\rm SUSY}\sim\mathcal{O}(10^3$~GeV), and we take the
interaction between the singlino and the MSSM fields to be of the
order of the EW scale, $v\sim\mathcal{O}(10^2$~GeV).

\section{The neutralino sector of the USSM}

We assume the MSSM gaugino unification relation
$M_1=(5/3)\tan^2\theta_W M_2 \approx 0.5 M_2$ and unified couplings
$g_X=g_Y$, but $M_1'$ will be taken as independent\footnote{For a
mechanism of generating non-universal U(1) gaugino masses, see e.g.
\cite{Suematsu:2006wh}.}  to investigate the impact of new states as a
function of $M_1'$. For the numerical values we take $M_2=1.5$ TeV,
$\mu=\lambda v_s/\sqrt{2}=0.3$ TeV, $m_s=g_X v_s=1.2$ TeV,
$\tan\beta=5$, $M_A=0.5$ TeV, neglect (small) $\tilde{B}$-$\tilde{B}'$
mixing, and adopt the E$_6$SSM assignment for the U$_X$(1)
charges~\cite{Choi:2006fz}.

Unlike the 4x4 MSSM case, the full 6x6 neutralino mass matrix cannot
be diagonalised analytically. However, since the mixing between the
new and MSSM states is small ${\mathcal O} (v)$ compared to $M_{\rm
SUSY}$, one can perform first the diagonalisation of the 4x4 MSSM and
the 2x2 $\tilde{S}$-$\tilde{B}'$ submatrices separately. Then the
perturbative expansion of the block-diagonalisation in $v/M_{\rm
SUSY}$ provides an excellent approximation to masses and mixings
\cite{Choi:2006fz}.

\begin{figure}[h!]
\begin{center}
\includegraphics[height=4.3cm,width=4.5cm,angle=0]{kalinowski_jan_fig1a.eps}
\includegraphics[height=4.5cm,width=4.cm,angle=0]{kalinowski_jan_fig1b.eps}\hspace{2mm}
\includegraphics[height=4.cm,width=3.5cm,angle=0]{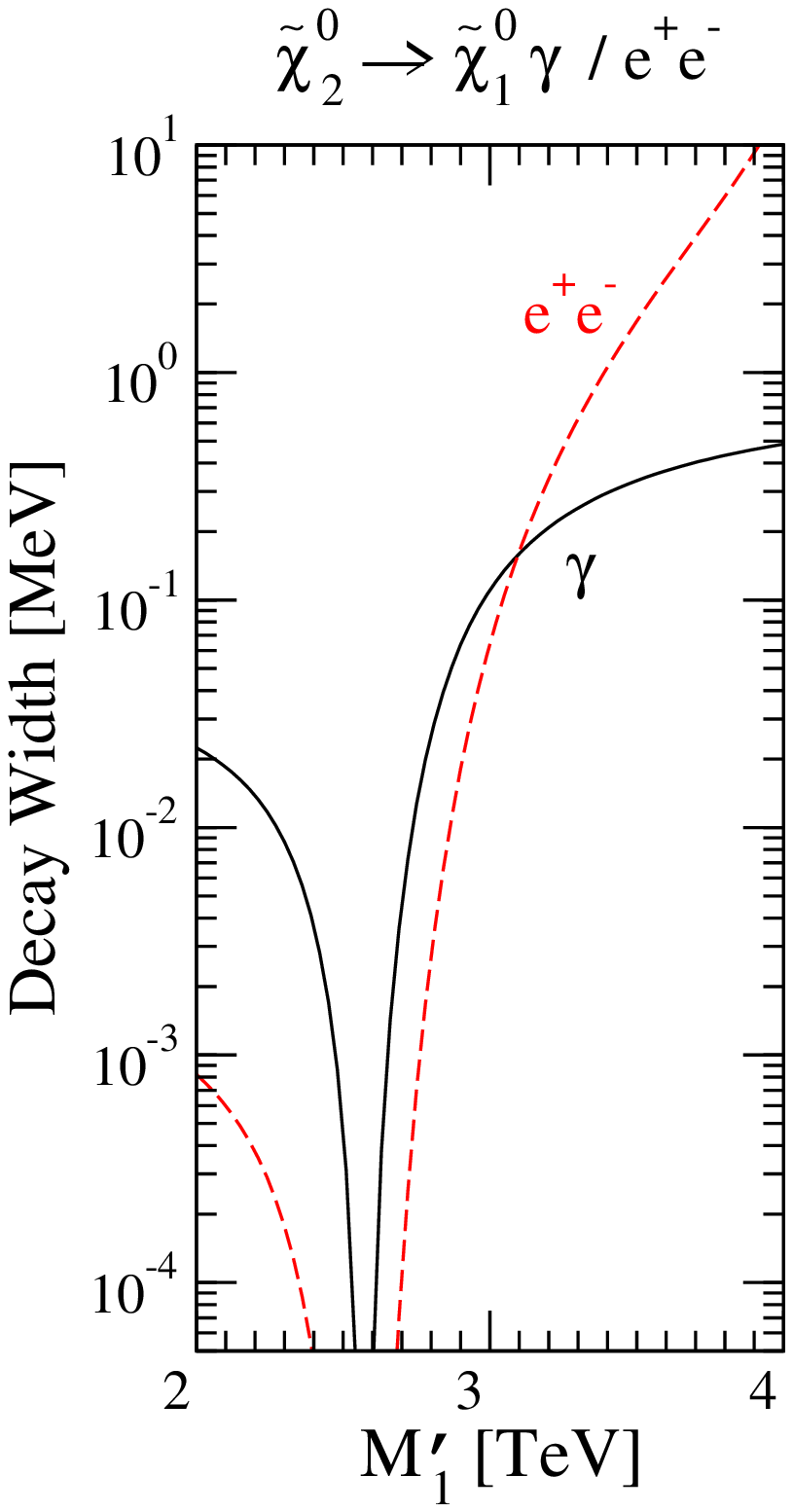}
\end{center}
\vskip -0.5cm \caption{\it The $M_1'$ evolution of (left) neutralino masses,
             (center) production cross sections
             for $\tilde{\chi}^0_1\tilde{\chi}^0_1$,
             $\tilde{\chi}^0_1\tilde{\chi}^0_2$ and
             $\tilde{\chi}^0_2\tilde{\chi}^0_2$
             pairs in $e^+e^-$ collisions, and (right) partial decay widths
             of $\tilde{\chi}^0_2$ (from \cite{Choi:2006fz}).
             }
\label{fig:kalinowski1}
\end{figure}

The mass spectrum is shown in Fig.\ref{fig:kalinowski1} (left) as a
function of $M_1'$.  For small $M'_1$ the eigenstates (denoted by
numbers with primes) are almost pure MSSM U(1) and SU(2) gauginos
$\tilde{\chi}^0_{1'},\tilde{\chi}^0_{2'}$, MSSM higgsinos
$\tilde{\chi}^0_{3'},\tilde{\chi}^0_{4'}$, and maximally mixed
U$_X$(1) gaugino and singlino states,
$\tilde{\chi}^0_{5'},\tilde{\chi}^0_{6'}$.  When $M'_1$ is shifted to
higher values, the mass eigenvalues in the new sector move apart,
generating strong cross--over patterns whenever a (signed) mass from
the new block comes close to one of the (signed) MSSM masses. This
happens at $M'_1 \approx 0.91$ TeV for $\tilde{\chi}^0_{6'}$ and
$\tilde{\chi}^0_{2'}$ states, and at $M'_1 \approx 2.68$ TeV for
$\tilde{\chi}^0_{4'}$ and $\tilde{\chi}^0_{5'}$. For higher $M_1'$ the
$\tilde{\chi}^0_{5'}$ approaches the singlino state and becomes the
LSP.

At an $e^+e^-$ collider the production processes $ e^+e^-\rightarrow
\tilde{\chi}^0_i\tilde{\chi}^0_j $ are generated by $s$--channel $Z_1$
and $Z_2$ exchanges (mass-eigenstates of $Z$ and $Z'$), and $t$-- and
$u$--channel $\tilde{e}_{L,R}$ exchanges.\footnote{ The numbering
without primes refers to mass eigenstates ordered according to
ascending masses.} In our scenario $M_{Z_2}=949$ GeV, the $ZZ'$ mixing
angle $\theta_{ZZ'}=3.3\times 10^{-3}$, and $m_{\tilde{e}_{R,L}}= 701$
GeV.  The $M_1'$ dependence of the production cross sections for the
three pairings of the two lightest neutralinos, $\{11\}, \{12\}$ and
$\{22\}$, is shown in Fig.\ref{fig:kalinowski1} (center) for
$\sqrt{s}=800$ GeV.  For small $M'_1$ the presence of $Z_2$ has little
influence on $\sigma\{\tilde{\chi}^0_1 \tilde{\chi}^0_2\}$ which is of
similar size as in the MSSM for mixed higgsino pairs.  However it
significantly enhances diagonal higgsino pairs
$\sigma\{\tilde{\chi}^0_1 \tilde{\chi}^0_1\}$ and
$\sigma\{\tilde{\chi}^0_2\tilde{\chi}^0_2\}$ compared with the MSSM,
even though the light neutralino masses are nearly identical in the
two models. At and beyond the cross--over with singlino, $M'_1 \approx
2.68$ TeV, dramatic changes set in for pairs involving the lightest
neutralino.

At the LHC the neutralinos will be analyzed primarily in cascade
decays of squarks or gluinos. In the USSM the cascade chains may be
extended compared with the MSSM by an additional step due to the
presence of two new neutralino states, for example, $ \tilde{u}_R\,
\to\, u \tilde{\chi}^0_6 \, \to\, u Z_1 \tilde{\chi}^0_5 \, \to\, u
Z_1 \ell \tilde{\ell}_R\, \to\, u Z_1 \ell \ell \tilde{\chi}^0_1, $
with partial decay widths significantly modified by the singlino and
bino' admixtures.  Also the presence of additional Higgs boson will
influence the decay chains.  Moreover, in the cross--over zones the
gaps between the masses of the eigenstates become very small
suppressing standard decay channels and, as a result, enhancing
radiative decays of neutralinos.  These decays are particularly
important in the cross--over at $M'_1 \simeq 2.6$, where the radiative
modes $ \tilde{\chi}^0_2, \tilde{\chi}^0_3 \to \tilde{\chi}^0_1 +
\gamma\,, \tilde{\chi}^0_3 \to \tilde{\chi}^0_2 + \gamma$ become
non--negligible, see Fig.\ref{fig:kalinowski1} (right).  Since the
photon will be very soft, these decays will be invisible making the
decay chains apparently shorter.

\section{USSM implications for dark matter}

\begin{figure}[h!]
\begin{center}
\includegraphics[height=4.5cm,width=7cm,angle=0,clip=]{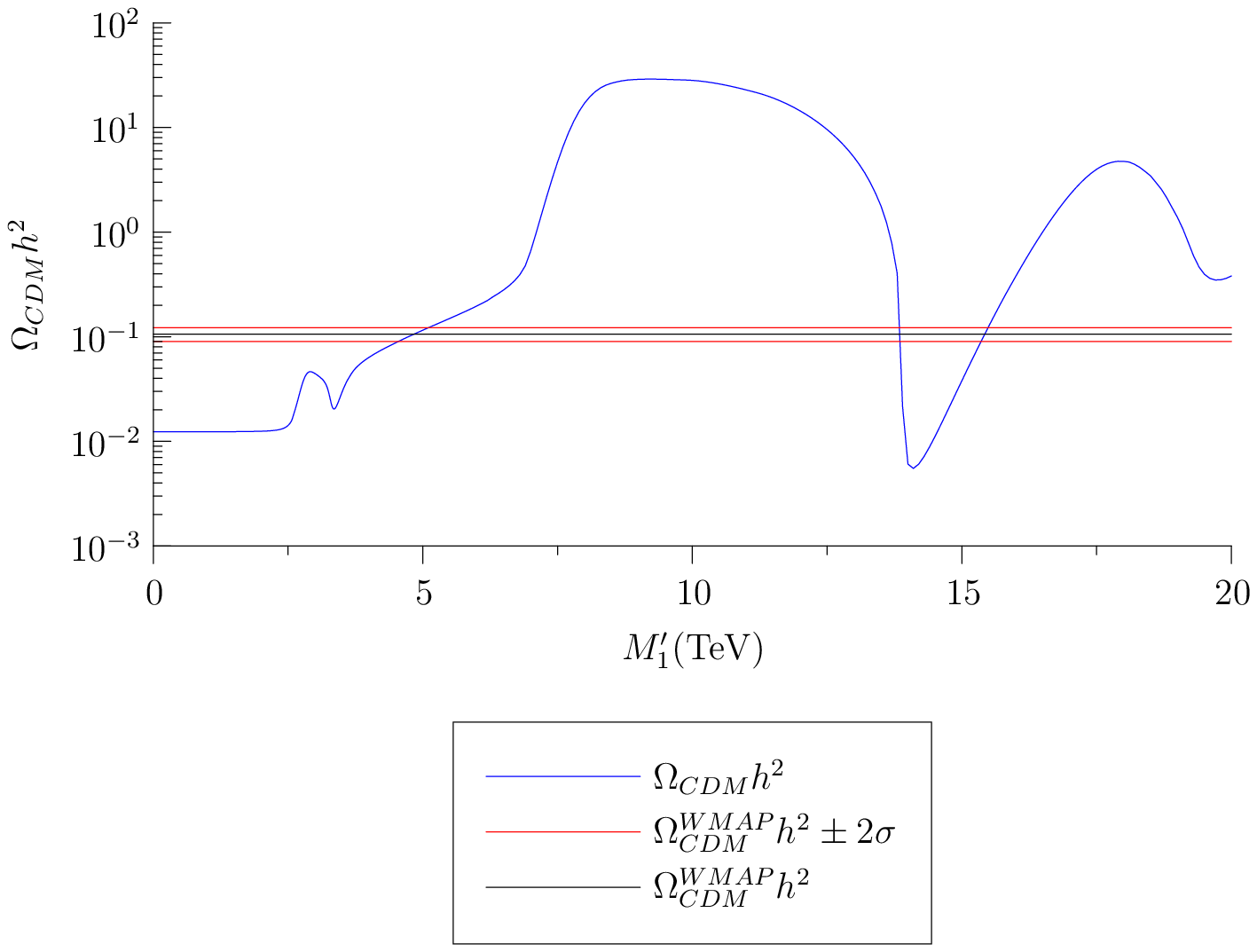}\hspace{5mm}
\includegraphics[height=5cm,width=6cm,angle=0]{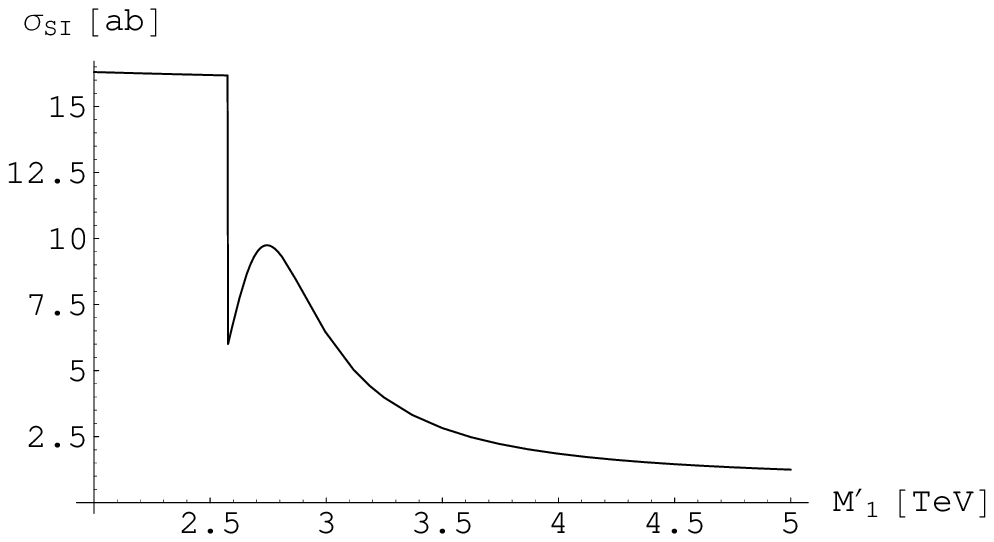}
\end{center}
\vskip -0.5cm \caption{\it The $M_1'$ dependence of (left) the
predicted relic density of DM, and (right) the elastic
spin-independent LSP-$^{73}$Ge cross section. We restrict the right
hand plot to 5 TeV as there are no new features above this energy.}
\label{fig:kalinowski2}
\end{figure}

If the lightest neutralino (LSP) is expected to be the source of the
relic abundance of dark matter in the universe, the predicted relic
density depends on the LSP composition. In the left panel of Fig.\ref{fig:kalinowski2} it is shown as a function of $M_1'$ \cite{jkkr}.
For small $M_1'$ the LSP is
almost an MSSM higgsino and for a mass $\sim 300$ GeV the predicted
value falls below the WMAP result.  As $M_1'$ increases, the singlino
admixture increases suppressing the LSP annihilation cross section and
the predicted relic density increases. The singlino LSP predominantly
annihilates via an off-shell s-channel singlet Higgs, which decays to
two light Higgs bosons. As $M_1'$ increases, the LSP mass decreases
and at $M'_1\approx 3.3$ TeV it reaches $m_{\tilde{\chi}^0_1}\approx
250$ GeV making the resonant annihilation via the heavy Higgs boson
efficient enough to lower the relic density. Further increase of
$M'_1$ switches off the heavy Higgs resonance and eventually the WMAP
value is met (shown as a horizontal band in Fig.\ref{fig:kalinowski2}
(left) \cite{jkkr}). Around $M'_1=7.5$ TeV the LSP becomes lighter
than the light Higgs. This switches off the annihilation via an
off-shell singlet Higgs, $\tilde{\chi}^0_1\tilde{\chi}^0_1\rightarrow
h_1 h_1$, normally the dominant annihilation mode of a singlino
LSP. As a result the relic density rises sharply. Further increasing
$M_1'$ decreases the LSP mass until it matches the resonant
annihilation channels of the light Higgs (at around 14 TeV) and $Z$
boson (at around 20 TeV). In both cases this results in a significant
dip in the relic density.

The singlino nature of the LSP is also of importance for direct DM searches. It has a strong impact on the elastic
spin-independent scattering off the nuclei, e.g. as shown in
Fig.\ref{fig:kalinowski2} (right)~\cite{jkkr} for the $^{73}$Ge
nucleus (the numerical codes have been developed in~\cite{dj}). For small
$M_1'$ the two lightest neutralinos (3' and 4' in
Fig.\ref{fig:kalinowski1}) are almost pure maximally mixed MSSM
higgsinos. When $M_1'$ increases, the mixing with singlino lowers
$m_{4'}$ so that at $M'_1\approx 2.6$ TeV the state 4' becomes the
LSP.  Since the higgsino mixing angles are such that the elastic
scattering of the state 4' is almost two orders of magnitude smaller
than for the state 3', it explains a sudden drop seen in
Fig.\ref{fig:kalinowski2} (right). At the same time the singlino and
bino' admixture of the LSP increases, which explains a local maximum
around 2.8 TeV. As the singlino component (the state 5') of the LSP
becomes dominant for higher $M_1'$ values, the elastic cross section
becomes smaller and smaller.

\section{Summary}
The U(1) extended MSSM provides an elegant way of solving the $\mu$
problem. As the neutralino sector is extended, the collider
phenomenology can significantly be altered and new scenarios for
matching the WMAP constraint can be realised. One example, in contrast
to the NMSSM, is that the USSM contains regions in which predominantly
singlino dark matter can fit the WMAP relic density measurement
without the need for coannihilation, or resonant s-channel
annihilation processes, where the LSP annihilates via
$\tilde{S}\tilde{B'}\rightarrow S^* \rightarrow hh$.


\section*{Acknowledgements}
Work supported by the Polish Ministry of Science and Higher Education
Grant No. 1 P03B 108 30 and the EC Project MTKDCT-2005-029466
"Particle Physics and Cosmology: the Interface". JPR would like to
thank Alexander Pukhov and Andre Semenov for useful advice.

\begin{footnotesize}


\end{footnotesize}

\begin{thebibliography}{99}
\bibitem{nmssm} For a recent summary and references, see D.J. Miller,
   R. Nevzorov and P.M. Zerwas, Nucl. Phys. B {\bf 681} (2004) 3.

\bibitem{ussm} D. Suematsu and Y. Yamagishi, Int. J. Mod. Phys. A
   {\bf 10} (1995) 4521; M. Cveti$\breve{\rm c}$, D.A. Demir, J.R. Espinosa,
   L.L. Everett and P. Langacker, Phys. Rev. D {\bf 56} (1997) 2861;
   {\bf 58} (1997) 119905(E).



\bibitem{King:2005jy}
  S.~F.~King, S.~Moretti and R.~Nevzorov,
  Phys.\ Rev.\  D {\bf 73} (2006) 035009
  [arXiv:hep-ph/0510419], and references therein.

\bibitem{Choi:2006fz}
  S.~Y.~Choi, H.~E.~Haber, J.~Kalinowski and P.~M.~Zerwas, Nucl.\ Phys.\ B
  {\bf 778} (2007) 85
  [arXiv:hep-ph/0612218], and references therein.

\bibitem{Barger:2007nv}
  V.~Barger, P.~Langacker, I.~Lewis, M.~McCaskey, G.~Shaughnessy and B.~Yencho,
  Phys.\ Rev.\  D {\bf 75} (2007) 115002
  [arXiv:hep-ph/0702036], and references therein.

\bibitem{jkkr} D. Jarecka, J. Kalinowski, S.F.King and J. Roberts, in preparation.




\bibitem{Suematsu:2006wh}
  D.~Suematsu,
  JHEP {\bf 0611} (2006) 029
  [arXiv:hep-ph/0606125].



\bibitem{dj} D. Jarecka, MSc thesis,
\verb$http://www.fuw.edu.pl/~djarecka/praca/praca11508dz.pdf$ (in Polish).



\end{thebibliography}
\end{document}